\documentclass{article}
\usepackage{amssymb} 
\usepackage{latexsym} 

\usepackage[french,english]{babel} 
\def\english{\selectlanguage{english}}
\def\french{\selectlanguage{french}}
\usepackage{RR}

\RRNo{4020}
\RRdate{October 2000} 
\RRauthor{Jan--Georg Smaus\thanks{ \texttt{jan.smaus@cwi.nl}, CWI, Kruislaan 413, 1098 SJ Amsterdam, The Netherlands.}
 \and Fran\c{c}ois Fages\thanks{\texttt{francois.fages@inria.fr}}
 \and Pierre Deransart\thanks{\texttt{pierre.deransart@inria.fr}}}

\authorhead{Smaus \& Fages \& Deransart} 
\RRtitle{Utilisation des modes pour garantir la propriété de ``subject reduction''
pour les programmes logiques typés avec sous-typage}

\RRetitle{Using Modes to Ensure Subject Reduction for Typed Logic Programs with Subtyping} 
\titlehead{Using modes to ensure subject reduction for typed logic programs} 
\RRnote{This paper is the complete version of a paper presented at
FST\&TCS 2000~\cite{SFD00}. It contains all proofs omitted there for
space reasons.
Copies of this report obtained from INRIA contain a mistake
(the assumptions of Lemma~\ref{more-specific-lemma} are incorrectly stated)
which is corrected in the present version.}

\RRresume{Nous considérons un système de types prescriptif avec polymorphisme para-métrique
et sous-typage pour les programmes logiques.
La propriété de ``subject reduction'' exprime la cohérence du système de types
vis à vis du modèle d'exécution: si un programme est ``bien typé'', alors
toutes les dérivations à partir d'un but ``bien typé'' sont encore
``bien typées''. Il est bien établi que sans sous-typage, cette propriété 
est vérifiée par les programmes logiques munis de leur modèle d'exécution
standard (non typé). Dans cet article nous donnons des conditions syntaxiques
qui garantissent cette propriété également en présence de relations
de sous-typage entre constructeurs de types. L'idée est de considérer
les programmes logiques ayant un flot de données fixe, déterminé par des modes.
}
\RRabstract{We consider a general
prescriptive type system with 
parametric polymorphism and subtyping for logic 
programs. 
The property of {\em subject reduction} expresses the consistency
of the type system w.r.t.\ the execution model:
if a program is ``well-typed'', 
then all derivations starting in a ``well-typed'' goal are again 
``well-typed''. 
It is well-established that without subtyping, this 
property is readily obtained for logic programs w.r.t.~their standard (untyped) 
execution model. Here we give syntactic conditions that ensure 
subject reduction also in the presence of general subtyping relations between 
type constructors. The idea is to consider 
logic programs with a fixed dataflow, given by modes. 
}
\RRmotcle{programmes logiques typés, modes, systèmes de types, sous-typage, ``subject reduction''}
\RRkeyword{typed logic programs, modes, type systems, subtyping, subject reduction}
\RRprojet{Contraintes} 
\RRtheme{2} 
\URRocq 

\def\onetom{\{1,\dots,m\}} 
\def\oneton{\{1,\dots,n\}} 
\def\zeroton{\{0,\dots,n\}} 
\newcommand{\Max}{{\mbox Max}} 
\newcommand{\vars}{{\mbox vars}} 
\newcommand{\pars}{{\mbox pars}} 
\newcommand{\restr}[1]{\!\!\upharpoonright_{#1}} 
\newcommand{\vect}[1]{{\bar{#1}}}  
\newcommand{\dom}{{ dom}} 
\newcommand{\ran}{{ ran}} 
\def\ftau{f_{\tau_1 \dots \tau_n\rightarrow \tau}} 
\def\ptau{p_{\tau_1 \dots \tau_n}} 
\newcommand{\inp}{{\it I}} 
\newcommand{\out}{{\it O}} 
\def\comment#1{\marginpar{\scriptsize #1}} 

\def\qed{}

\begin{document}
\bibliographystyle{plain}

\makeRR   

 
\newtheorem{theorem}{Theorem}{\bfseries}{\rm} 
\newtheorem{rmtheorem}[theorem]{Theorem}{\bfseries}{\rm} 
\newtheorem{rmlemma}[theorem]{Lemma}{\bfseries}{\rm} 
\newtheorem{coro}[theorem]{Corollary}{\bfseries}{\rm} 
\newtheorem{propo}[theorem]{Proposition}{\bfseries}{\rm} 
\newtheorem{definition}{Definition}{\bfseries}{\rm} 
\newtheorem{defi}[definition]{Definition}{\bfseries}{\rm} 
\newtheorem{example}{Example}{\bfseries}{\rm} 
\newtheorem{ex}[example]{Example}{\bfseries}{\rm} 

\newenvironment{proof}{\bfseries Proof:\rm} {\hfill$\square$\vspace{4mm}}

\section{Introduction}\label{intro-sec} 
Prescriptive types are used in logic and functional programming  
to restrict the underlying syntax so that only ``meaningful'' 
expressions are allowed. This allows for many programming errors to be 
detected by the compiler. G\"odel~\cite{goedel} and
Mercury~\cite{mercury} are two implemented typed logic programming
languages. 

A natural stability property one desires for a type 
system is that it is consistent with the execution model: 
once a program has passed the compiler,  
it is guaranteed that ``well-typed'' configurations will only generate 
``well-typed'' configurations at runtime. 
Adopting the terminology from the theory of 
the \mbox{$\lambda$-calculus~\cite{Tho91}}, this property of a typed program 
is called {\em subject reduction}. For the simply typed 
$\lambda$-calculus, subject reduction states that the type of a 
$\lambda$-term is invariant under reduction. This translates in a 
well-defined sense to functional and logic programming.  
 
Semantically, a type represents a set of  
terms/expressions~\cite{HT92-new,LR91}.  
Now subtyping makes type systems more expressive and flexible in that 
it allows to express inclusions among these sets. For example, 
if we have types $\mathit{int}$ and $\mathit{real}$ defined in the usual way, we would 
probably want to declare $\mathit{int} \leq \mathit{real}$, i.e., 
 the set of integers is a subset of the set of reals.  
More generally, subtype relations like for example $\mathit{list(u)}<\mathit{term}$, 
which expresses the possibility of viewing a list as a term, make it 
possible to type Prolog meta-programming predicates \cite{FP98},
as shown in Ex.~\ref{defend-against-DH88-ex} below and 
Sec.\ \ref{discussion-sec}. 

In functional programming, a type system that includes subtyping 
would then state that wherever an expression of type $\sigma$ is  
expected as an argument, any expression having a type  
$\sigma'\leq\sigma$ may occur. Put differently, an expression of type  
$\sigma$ can be used wherever an expression of type 
$\sigma'\geq\sigma$ is expected. The following example explains this 
informally, using an ad hoc syntax. 
 
 \begin{example}\label{sqrt-ex} 
Suppose we have two functions $sqrt: \mathit{real}\rightarrow \mathit{real}$ and  
$\mathit{fact}: int\rightarrow int$ which compute the square root and 
factorial, respectively.  Then $sqrt\ (\mathit{fact}\ 3)$ is a legal 
expression, since $\mathit{fact}\ 3$ is of type $\mathit{int}$ and may 
therefore be used as an argument to $sqrt$, because $sqrt$ expects an 
argument of type $\mathit{real}$, and $\mathit{int}<\mathit{real}$. 
\end{example}

Subject reduction in functional programming crucially relies on the 
fact that there is a clear notion of dataflow. It is always the  
{\em arguments} (the ``input'') of a function that may be smaller 
than expected, and the result (the ``output'') may be greater 
than expected. This is best illustrated by a counterexample, which is 
obtained by introducing {\em reference} types. 
 
\begin{example}\label{counterex-ex} 
Suppose we have a function  
\[ 
\begin{array}{l} 
f:\ \mathit{real} \ \mathit{REF} \rightarrow \mathit{real}\\ 
\mathit{let}\ f(x)=\; x:=3.14;\ \mathit{return}\ x 
\end{array} 
\]

So $f$ takes a {\em reference} (pointer) to a real as argument, 
assigns the value $3.14$ to this real, and also return $3.14$.  
Even though $\mathit{int}<\mathit{real}$, this function cannot be applied to an  
$\mathit{int}\ \mathit{REF}$, since the value $3.14$ cannot be assigned to an integer.  
\end{example}

In the example,  
the variable $x$ is used both for input and output, and hence there is 
no clear direction of dataflow. While this problem 
is marginal in functional programming (since reference types play no 
essential role in the paradigm), it is the main problem for subject 
reduction in logic programming with subtypes, as we show in the next
example.

 
Subject reduction for logic programming means that resolving a 
``well-typed'' goal with a ``well-typed'' clause will always result in 
a ``well-typed'' goal. It holds for parametric 
polymorphic type systems without
subtyping~\cite{LR91,MO84}.\footnote{
However, it has been pointed out~\cite{Han89,HT92-new} that the first 
formulation of subject reduction by Mycroft and O'Keefe~\cite{MO84} 
was incorrect, namely in ignoring the transparency condition, which we 
will define in Section~\ref{type-system-sec}.}

\begin{example}\label{sqrt-logic-ex} 
In analogy to Ex.~\ref{sqrt-ex}, suppose $\mathtt{Sqrt}/2$ and 
$\mathtt{Fact}/2$ are predicates of declared type 
$\tt (Real, Real)$ and $\tt (Int, Int)$, respectively. 
Consider the program 
\begin{verbatim} 
Fact(3,6). 
Sqrt(6,2.449). 
\end{verbatim} 
 
and the derivations 
\[ 
\begin{array}{l} 
\tt  
Fact(3,x),\ Sqrt(x,y) \leadsto 
Sqrt(6,y) \leadsto \Box\\ 
\tt Sqrt(6,x),\ Fact(x,y) \leadsto 
Fact(2.449,y) 
\end{array} 
\] 
In the first derivation, all arguments always have a type that is 
less than or equal to the declared type, and so we have 
subject reduction. In the second derivation, the argument  
$\tt 2.449$ to $\tt Fact$ has type $\tt Real$, which is strictly 
greater than the declared type. The atom $\tt Fact(2.449,y)$ is 
illegal, and so we do not have  
subject reduction. 
\end{example}

In this paper, we address this problem by giving a fixed direction of 
dataflow to logic programs. This is done by introducing 
modes~\cite{A97} and replacing unification with double 
matching~\cite{AE93}, so that the dataflow is always from the input to
the output positions in an atom. 
We impose a condition on the types of
terms in the output positions, or more precisely, on the types of the
{\em variables} occurring in these terms: each variable must have
{\em exactly} the declared (expected) type of the position where it occurs.
 
In Ex.\ \ref{sqrt-logic-ex}, let the first argument of each 
predicate be input and the second be output. In both derivations, 
$\tt x$ has type $\tt Int$. For the atom 
$\tt Fact(3,x)$, this is exactly the declared type, and so the 
condition is fulfilled for the first derivation. 
In contrast, for the atom 
$\tt Sqrt(6,x)$, the declared type is $\tt Real$, and so the condition 
is violated.

The contribution of this paper is a statement that programs that 
are typed according to a type system with subtyping, and
respect certain conditions concerning the modes,
enjoy the subject reduction property, i.e., the type system
is consistent w.r.t.~the (untyped) execution model.
This means that 
effectively the types can be ignored at runtime, which has
traditionally been considered as desirable, although there are also
reasons for keeping the types during execution~\cite{NP92}.
In Sec.~\ref{discussion-sec}, we discuss the conditions on programs. 

Most type systems with subtyping for logic programming languages 
that have been proposed 
are descriptive type systems, i.e.~their purpose is to describe 
the set of terms for which a predicate is true. 
There are few works considering prescriptive type systems for  
logic programs with 
subtyping~\cite{B95,DH88,FP98,Han92,HT92-new}.  
Hill and Topor~\cite{HT92-new} give a result on subject reduction only for 
systems without subtyping, and study general type systems with subtypes. 
However their results on the existence of principal typings 
for logic programs with subtyping turned out to be  
wrong, as pointed out by Beierle~\cite{B95}. 
He shows the existence of principal typings with subtype relations 
between constant types only, and provides type inference algorithms. 
Beierle and also 
Hanus~\cite{Han92} do not claim subject reduction for the systems they 
propose. Fages and Paltrinieri~\cite{FP98} have 
shown a weak form of subject reduction for constraint logic programs 
with general subtyping relations, where equality constraints
replace term substitutions in the execution model. 
 
On the other hand, the idea of introducing modes to ensure subject reduction for  
standard logic programs was already proposed by Dietrich and 
Hagl~\cite{DH88}.  
However they do not study the decidability of  
the conditions they impose on the subtyping relation. Furthermore since each result 
type must be transparent (a condition we will define later), this 
means effectively that in general,  
subtype relations between type 
constructors of different arities are forbidden. We illustrate this 
with an example. 
 
\begin{example}\label{defend-against-DH88-ex} 
Assume types $\tt Int$, $\tt String$ and $\mathtt{List}(\mathtt{u})$ 
defined as usual, and a type $\tt Term$ that contains all terms  
(so all 
types are subtypes of $\tt Term$). Moreover, assume $\tt Append$ 
as usual with declared type $\tt (List(u),List(u),List(u))$, 
and a predicate $\tt Functor$ with declared type $\tt (Term,String)$, 
which gives the top functor of a term. In our formalism, we could 
show subject reduction for the query  
$\tt Append([1],[],x),\ Functor(x,y)$, 
whereas this is not possible in~\cite{DH88} because the subtype relation 
between $\tt List(Int)$ and $\tt Term$ cannot be expressed. 
\end{example}

The plan of the paper is as follows. 
Section~\ref{type-system-sec} mainly introduces 
the type system. In Sec.~\ref{hierarchy-sec}, we show how 
expressions can be typed assigning different types to the variables, 
and we introduce {\em ordered substitutions}, which are substitutions  
preserving types, and thus ensuring subject reduction. In  
Sec.~\ref{max-types-sec}, we show under which conditions substitutions 
obtained by unification are indeed ordered. In 
Sec.~\ref{nicely-typed-sec}, we show how these conditions on unified 
terms can be translated into conditions on programs and derivations.

\section{The Type System}\label{type-system-sec}\label{prelim-sec} 
We will use the type system of \cite{FP98}. 
First we recall some basic concepts~\cite{A97}.  When we refer to a {\em clause in a program}, we mean a copy 
of this clause whose variables are renamed apart from variables 
occurring in other objects in the context. A query is a sequence of 
atoms. A query is a sequence of 
atoms. A query $Q'$ is a {\bf resolvent of} a query $Q$ and a clause
$H\leftarrow \mathbf{B}$ if
$Q= A_1,\dots,A_m$,  
$Q'=(A_1,\dots,A_{k-1},\mathbf{B},A_{k+1},\dots,A_m)\theta$,  
and $H$ and $A_k$ are unifiable with MGU $\theta$. 
{\bf Resolution steps} and 
{\bf derivations} are defined in the usual way.

\subsection{Type expressions}\label{program-symbols-subsec} 
The set of types $\mathcal T$ is given by the term structure based on 
a finite set of {\bf constructors} $\mathcal K$, where with each 
$K\in\mathcal{K}$ an arity 
$m\geq 0$ is associated (by writing $K/m$), 
and a denumerable set $\mathcal U$ of 
{\bf parameters}. A {\bf flat type} is a type of the form  
$K(u_1,\dots,u_m)$, where $K \in \mathcal K$ and the $u_i$ are  
distinct parameters. 
We write $\tau[\sigma]$ to denote that the type $\tau$ 
strictly contains the type 
$\sigma$ as a subexpression. 
We write $\tau[u/\sigma]$ to denote the type obtained by replacing
all the occurrences of $u$ by $\sigma$ in $\tau$.
The size of a type $\tau$, defined as the number of occurrences of constructors and parameters in $\tau$,
is denoted by $\mbox{size}(\tau)$.
 
A {\bf type substitution} $\Theta$ is an idempotent mapping 
from parameters to types that is the identity almost everywhere. 
{\bf Applications} of type substitutions are defined 
in the obvious way. 
The {\bf domain} of a type substitution is
denoted by $\dom$, the parameters in its {\bf range} by $\ran$.
The set of parameters in a syntactic object $o$ is 
denoted by $\pars(o)$. 

We now qualify what kind of subtyping we allow. Intuitively, when a
type $\sigma$ is a subtype of a type $\tau$, this means that each term
in $\sigma$ is also a term in $\tau$. The subtyping relation $\leq$ is
designed to have certain nice algebraic properties, stated in
propositions below.

We assume an 
order $\leq$ on type constructors 
such that: 
$K/m\leq K'/m'$ implies $m\geq m'$;
and, for each $K\in\mathcal K$, the set  
$\{K' \mid K\leq K'\}$ has a maximum. 
Moreover, we assume that with each pair 
$K/m\leq K'/m'$, an injection 
$\iota_{K,K'}: \{1,\dots,m'\} \rightarrow \onetom$ 
is associated such that $\iota_{K,K''}=\iota_{K,K'}\circ\iota_{K',K''}$
whenever $K\le K'\le K''$.
This order is extended to the {\bf subtyping order} on
types, denoted by $\leq$, as the least relation satisfying the rules in 
Table \ref{subtyping-tab}. 

\begin{table}[ht]
\begin{center}
\begin{tabular}{lll}
{\em (Par)} & 
$u\leq u$ & 
$u$ is a parameter\\[2ex]
{\em (Constr)} &
\Large
$\frac%
  {\tau_{\iota(1)}\leq\tau'_1\ \dots \ \tau_{\iota(m')}\leq\tau'_{m'}}%
  {K(\tau_1,\dots,\tau_m)\leq K'(\tau'_1,\dots,\tau'_{m'})}\quad$ &
$K\leq K'$, $\iota = \iota_{K, K'}$.
\end{tabular}
\end{center}
\caption{The subtyping order on types\label{subtyping-tab}}
\end{table}

\begin{propo}\label{size-prop}
If $\sigma\leq\tau$ then $\mbox{size}(\sigma)\geq \mbox{size}(\tau)$ .
\end{propo}
\begin{proof}
By structural induction on $\tau$.
\qed\end{proof}

\begin{propo}\label{inst-closed-prop}
If $\sigma\leq\tau$ then $\sigma\Theta\leq\tau\Theta$ for any type substitution $\Theta$.
\end{propo}
\begin{proof}
By structural induction on $\tau$.
\qed\end{proof}

\begin{propo}\label{max-exists} 
For each type $\tau$, the set
$\{\sigma \mid \tau\leq\sigma\}$ has a maximum, which is denoted by 
$\Max(\tau)$.
\end{propo}
\begin{proof}
By structural induction on $\tau$.
\qed\end{proof}

\begin{propo}\label{max-prop} 
For 
all types $\tau$ and $\sigma$, $\Max(\tau[u/\sigma])=\Max(\tau)[u/\Max(\sigma)]$.
\end{propo}
\begin{proof}
By structural induction on $\tau$.
\qed\end{proof}

Note that for  Prop.\ \ref{max-exists}, it is crucial that we require
that $K/m\leq K'/m'$ implies $m\geq m'$, that is, as we move up in the subtype hierarchy, the arity of the type
constructors does not increase. For example, if we allowed for 
$\tt Emptylist/0 \leq List/1$, then by Prop.~\ref{inst-closed-prop}, 
we would also have $\tt Emptylist \leq List(\tau)$ for all types
$\tau$, and so, Prop.~\ref{max-exists} would not hold. 
Note that the possibility of ``forgetting'' type parameters in
subtype relations, as in $\tt List/1\le Anylist/0$, may provide solutions
to inequalities of the form $\tt List(u)\le u$, e.g.~$\tt u=Anylist$.
However, we have:

\begin{propo}\label{propfini}\label{corfini}
An inequality of the form $u\le \tau[u]$ has no solution.
An inequality of the form $\tau[u]\le u$ has no solution if $u\in pars(\Max(\tau))$.
\end{propo}
\begin{proof}
For any type $\sigma$, we have $\mbox{size}(\sigma)<\mbox{size}(\tau[\sigma])$,
hence by Prop~\ref{size-prop}, $\sigma\not\le\tau[\sigma]$,
that is $u\le \tau[u]$ has no solution.

For the second proposition, we prove its contrapositive.
Suppose $\tau[u]\le u$ has a solution, say $\tau[u/\sigma]\le \sigma$.
By definition of a maximum and Prop.~\ref{max-exists}, we have $\Max(\sigma)=\Max(\tau[u/\sigma])$.
Hence by Prop.~\ref{max-prop}, $\Max(\sigma)=\Max(\tau)[u/\Max(\sigma)]$.
By the rules in Table \ref{subtyping-tab}, 
$u\neq\Max(\tau)$. Therefore $u\not\in\pars(\Max(\tau))$, 
since otherwise $\Max(\sigma)=\Max(\tau)[u/\Max(\sigma)]$ would contain 
$\Max(\sigma)$ as a strict subexpression which is impossible.
\qed\end{proof}

\subsection{Typed programs}\label{typed-language-subsec}
We assume a denumerable set $\mathcal V$ of {\bf variables}.
The set of variables in a syntactic object $o$ is
denoted by $\vars(o)$.   
We assume a finite set $\mathcal F$ (resp.~$\mathcal P$) 
of {\bf function} (resp.~{\bf predicate}) symbols, each
with an arity and a {\bf declared type} associated with it, 
such that:
for each $f \in \mathcal F$, the declared type has the form
$(\tau_1,\dots,\tau_n,\tau)$, 
where $n$ is the arity of $f$,
$(\tau_1,\dots,\tau_n)\in {\mathcal T}^n$,
$\tau$ is a flat type
and satisfies the {\em transparency condition} \cite{HT92-new}:\label{transparency}
$\pars(\tau_1,\dots,\tau_n) \subseteq \pars(\tau)$;
for each $p \in \mathcal P$, the declared type has the form
$(\tau_1,\dots,\tau_n)$, 
where $n$ is the arity of $p$ and
$(\tau_1,\dots,\tau_n)\in {\mathcal T}^n$. 
The declared types are indicated by writing 
$\ftau$ and $\ptau$, however it is assumed that the parameters in
$\tau_1,\dots,\tau_n,\tau$ are fresh for each occurrence of $f$ or $p$.
We assume that there is a special predicate symbol 
$=_{\tt u,u}$
where ${\tt u}\in \mathcal U$.

Throughout this paper, 
we assume that $\mathcal K$, $\mathcal F$, and 
$\mathcal P$ are fixed by means of declarations in a 
{\bf typed program}, where the syntactical
details are insignificant for our results.
In examples we loosely follow G\"odel syntax~\cite{goedel}. 

A {\bf variable typing} (also called {\em type context}~\cite{FP98})
is a mapping from a finite subset of $\mathcal V$ to $\mathcal T$,
written as $\{x_1:\tau_1,\dots,x_n:\tau_n\}$. The restriction of a 
variable typing $U$ to the variables in a
syntactic object $o$ is denoted as $U \restr{o}$. 
The type system, which defines terms, atoms etc.\ relative to a variable
typing $U$, consists of the rules shown in Table \ref{rules-tab}.    

\begin{table}[ht]
\begin{center}
\begin{tabular}{lll}
{\em (Var)} &
$\{x:\tau,\dots\}\vdash x:\tau$\\[2ex]
{\em (Func)} &
\Large
$\frac%
  {U\vdash t_i:\sigma_i \ \sigma_i\leq\tau_i\Theta \ (i\in\oneton)}%
  {U\vdash\ftau(t_1,\dots,t_n):\tau\Theta}$ &
$\Theta$ is a type substitution\\[2ex]
{\em (Atom)} &
\Large
$\frac%
  {U\vdash t_i:\sigma_i \ \sigma_i\leq\tau_i\Theta \ (i\in\oneton)}%
  {U\vdash\ptau(t_1,\dots,t_n) \mathit{Atom}}$ &
$\Theta$ is a type substitution\\[2ex]
{\em (Headatom)} &
\Large
$\frac%
  {U\vdash t_i:\sigma_i \ \sigma_i\leq\tau_i \ (i\in\oneton)}%
  {U\vdash\ptau(t_1,\dots,t_n) \mathit{Headatom}}$\\[2ex]
{\em (Query)} &
\Large
$\frac%
  {U\vdash A_1\ \mathit{Atom}\ \dots \  U\vdash A_n\ \mathit{Atom}}%
  {U\vdash A_1,\dots,A_n\ \mathit{Query}}$ \\[2ex]
{\em (Clause)} & 
\Large
$\frac%
  {U\vdash Q\ Query \quad U\vdash A\ \mathit{Headatom}}%
  {U \vdash A \leftarrow Q\ \mathit{Clause}}$
\end{tabular}
\end{center}
\caption{The type system.\label{rules-tab}}
\end{table}

If for an object, say a term $t$, we can deduce for some variable
typing $U$ and some type $\tau$ that $U\vdash t:\tau$, intuitively
this term is {\em well-typed}. Otherwise the term is {\em ill-typed}
(and likewise for atoms, etc.).

\section{The Subtype and Instantiation Hierarchies}\label{hierarchy-sec}

\subsection{Modifying Variable Typings}
Here we present the following result: if we can derive
that some object is in the typed language using a variable typing
$U$, then we can always modify $U$ in three ways: extending its
domain, instantiating the types, and making the types smaller. First
we define:

\begin{defi}\label{var-typing-order-def}
Let $U$, $U'$ be variable typings. We say that $U$ is {\bf smaller or equal}
$U'$, denoted $U \leq U'$, if 
$U=\{x_1:\tau_1,\dots,x_n:\tau_n\}$, 
$U'=\{x_1:\tau'_1,\dots,x_n:\tau'_n\}$, and for all $i\in\oneton$, we
have $\tau_i \leq \tau'_i$.

The symbols $<$, $\geq$, $>$ are defined in the obvious way.
\end{defi}

We use the notation $U' \supseteq \leq U$, which means that there
exists a variable typing $U''$ such that $U' \supseteq U''$ and $U''
\leq U$.

\begin{rmlemma}
\label{variable-typing-lemma}
Let $U$, $U'$ be variable typings and $\Theta$ a type
substitution such that $U' \supseteq \leq U\Theta$. 
If $U\vdash t:\sigma$, then 
$U'\vdash t:\sigma'$ where $\sigma' \leq \sigma\Theta$.
Moreover, if $U\vdash A\ \mathit{Atom}$ then $U'\vdash A\ \mathit{Atom}$, and if
$U\vdash Q\ Query$ then $U'\vdash Q\ Query$. 
\end{rmlemma}
\begin{proof}
The proof of the first part is by structural induction.  For the base
case, suppose $t\in \mathcal V$. Then by Rule {\em (Var)}, 
$t:\sigma \in U$ and hence for some $\sigma'\leq\sigma\Theta$, we have
$t:\sigma' \in U'$. Thus again by {\em (Var)}, $U'\vdash t:\sigma'$.

Now consider the case $t=\ftau(t_1,\dots,t_n)$ 
where the inductive hypothesis holds for $t_1,\dots,t_n$. 
By Rule {\em (Func)}, there exists a type 
substitution $\Theta'$ such that 
$\tau\Theta'=\sigma$,
and $U\vdash t_i:\sigma_i$ where
$\sigma_i\leq\tau_i\Theta'$ for each $i \in \oneton$. 
Thus by Prop.~\ref{inst-closed-prop}, 
$\sigma_i\Theta\leq\tau_i\Theta'\Theta$.
By the inductive hypothesis, for all $i\in\oneton$ we have
$U'\vdash t_i: \sigma'_i$ where
$\sigma'_i\leq\sigma_i\Theta$, therefore by transitivity of $\leq$ we
have $\sigma'_i\leq\tau_i\Theta'\Theta$ and hence by Rule 
{\em (Func)},
$U'\vdash t:\tau\Theta'\Theta$ 
(i.e.\ $U'\vdash t:\sigma\Theta$).

Now suppose $A=\ptau(t_1,\dots,t_n)$. 
By Rule {\em (Pred)}, there exists a type substitution $\Theta'$ such
that $U\vdash t_i:\sigma_i$ where $\sigma_i\leq\tau_i\Theta'$ for each
$i \in \oneton$.  Thus by Prop.~\ref{inst-closed-prop},
$\sigma_i\Theta\leq\tau_i\Theta'\Theta$.  
By the first part of the statement, for all $i\in\oneton$ we have
$U'\vdash t_i: \sigma'_i$ where $\sigma'_i\leq\sigma_i\Theta$,
therefore by transitivity of $\leq$ we have
$\sigma'_i\leq\tau_i\Theta'\Theta$ and hence by Rule {\em (Pred)}, 
$U'\vdash A\ \mathit{Atom}$.

The final case for a query follows directly from Rule {\em (Query)}. 
\qed\end{proof}

\subsection{Typed Substitutions}
Typed substitutions are a fundamental concept for typed logic
programs. Ignoring subtyping for the moment, a typed substitution
replaces each variable with a term of the same type as the variable.

\begin{defi}\label{substitution-def}
If 
$U\vdash x_1\!=\!t_1, \dots, x_n\! =\! t_n\ Query$
where $x_1,\dots,x_n$ are distinct variables and for
each $i\in \oneton$, $t_i$ is a term distinct from $x_i$, then 
$(\{ x_1/t_1,\dots,x_n/t_n\}, U)$ is a 
{\bf typed (term) substitution}. The application of a substitution is
defined in the usual way.
\end{defi}

To show that applying a typed
substitution preserves ``well-typedness'' for systems with subtyping,
we need a further condition.  Given a typed substitution 
$(\theta,U)$, the type assigned to a variable $x$ by $U$ must be
sufficiently big, so that it is compatible with the type of the term
replaced for $x$ by $\theta$. 

\begin{example}\label{unordered-ex}
Consider again Ex.~\ref{sqrt-logic-ex}. 
As expected, assume that $\tt 3, 6$ have declared type
$\tt Int$, and $\tt 2.449$ has declared type $\tt Real$, and 
$\tt Int \leq Real$. Given the 
variable typing
$U=\{\tt x:Int, y:Int\}$, we have
$U\vdash {\tt x: Int}$, 
$U\vdash {\tt 2.449: Real}$, and hence 
$U\vdash {\tt x=2.449}\ \mathit{Atom}$. So 
$(\{{\tt x/2.449}\},U)$ is a typed substitution. Now we have 
$U\vdash {\tt Fact(x,y)}\ \mathit{Atom}$, but 
$U\not\vdash {\tt Fact(2.449,y)}\ \mathit{Atom}$.
\end{example}


In the previous example, the type of $\tt x$ is too small to
accommodate for instantiation to $\tt 2.449$. This motivates the 
following definition.

\begin{defi}\label{ordered-subst-def}
A typed (term) substitution $(\{x_1/t_1,\dots,x_n/t_n\},U)$ is an 
{\bf ordered substitution} if, for each $i \in \oneton$, where
$x_i:\tau_i \in U$, there exists $\sigma_i$ such that 
$U\vdash t_i:\sigma_i$ and $\sigma_i \leq \tau_i$.
\end{defi}

The following result states that expressions stay ``well-typed'' when 
ordered substitutions are
applied~\cite[Lemma~1.4.2]{HT92-new}. Moreover, the type of terms may
become smaller.

\begin{rmlemma}\label{apply-substitution-lemma}
Let $(\theta,U)$ be an ordered substitution.
If $U\vdash t:\sigma$ then $U\vdash t\theta :\sigma'$ for some
$\sigma'\leq\sigma$. Moreover, if 
$U\vdash A\ \mathit{Atom}$ then $U\vdash A\theta\ \mathit{Atom}$,
and likewise for queries and clauses.
\end{rmlemma}
\begin{proof}
The proof of the first part is by structural induction. For the base
case, suppose $t\in \mathcal V$. Then by Rule {\em (Var)}, 
$t:\sigma \in U$. If $t\theta=t$, there is nothing to show. If 
$t/s \in \theta$, then by definition of an ordered substitution,
$U\vdash s:\sigma'$ and hence 
$U\vdash t \theta:\sigma'$ where $\sigma'\leq\sigma$.

Now consider the case $t=\ftau(t_1,\dots,t_n)$ 
where the inductive hypothesis holds for $t_1,\dots,t_n$. 
By Rule {\em (Func)}, there exists a type
substitution $\Theta$ such that $\tau\Theta=\sigma$, 
and $U\vdash t_i:\sigma_i$ where
$\sigma_i\leq\tau_i\Theta$ for each $i \in \oneton$. 
By the inductive hypothesis, 
for all $i\in\oneton$ we have
$U\vdash t_i\theta: \sigma'_i$ where
$\sigma'_i\leq\sigma_i$, 
and hence by transitivity of $\leq$ and Rule {\em (Func)}, 
$U\vdash t:\sigma$ (i.e.~$\sigma'=\sigma$).

Now consider an atom $A=\ptau(t_1,\dots,t_n)$. 
By Rule {\em (Pred)}, there exists a type
substitution $\Theta$ such that 
such
that $U\vdash t_i:\sigma_i$ where 
$\sigma_i\leq\tau_i\Theta$ for each
$i \in \oneton$.
By the inductive hypothesis, 
for all $i\in\oneton$ we have 
$U\vdash t_i\theta:\sigma'_i$ where
$\sigma'_i \leq \sigma_i$, and hence by Rule
{\em (Atom)}, 
$U\vdash A\theta\ \mathit{Atom}$.
\qed\end{proof}

\section{Conditions for Ensuring Ordered Substitutions}\label{max-types-sec}

In this section, we show under which conditions it can be
guaranteed that the substitutions applied in resolution steps are
ordered substitutions.

\subsection{Type Inequality Systems}
The substitution of a resolution step is obtained by unifying two
terms, say $t_1$ and $t_2$. In order for the
substitution to be typed, it is necessary that we can derive
$U\vdash t_1=t_2\ \mathit{Atom}$ for some variable typing $U$. We will show that if 
$U$ is, in a certain sense, maximal, then it
is guaranteed that the typed substitution is ordered. 

We need to formalise a straightforward concept, namely paths
leading to subterms of a term.

\begin{defi}\label{position-def}
A term $t$ has the subterm $t$ in {\bf position} $\epsilon$.
If $t = f(t_1,\dots,t_n)$ and $t_i$ has subterm $s$ in position
$\zeta$, then $t$ has subterm $s$ in position $i.\zeta$.
\end{defi}

\begin{example}
The term $\tt F(G(C),H(C))$ has subterm $\tt C$ in position $1.1$, but
also in position $2.1$. The position $2.1.1$ is undefined for this term. 
\end{example}

Let us use the notation $\_ \vdash t:\leq \sigma$ as
a shorthand for: there exists a variable typing $U$ and a 
type $\sigma'$ such that  
$U\vdash t:\sigma'$ and $\sigma'\leq\sigma$.
To derive $U\vdash t_1=t_2\ \mathit{Atom}$, it is clear that the last step has the
form
\[
\frac%
  {U\vdash t_1:\tau_1 \quad U\vdash t_2:\tau_2 \quad \tau_1\leq{\tt u}\Theta \quad \tau_2\leq{\tt u}\Theta}%
  {U\vdash t_1 =_{{\tt u,u}} t_2\ \mathit{Atom}} 
\]

That is to say, we use an {\em instance} 
$({\tt u,u})\Theta$ of the declared type of the equality
predicate, and the types of $t_1$ and $t_2$ are both less than or
equal to ${\tt u}\Theta$.  This motivates the following question: Given a
term $t$ such that $\_ \vdash t:\leq \sigma$, what are the maximal
types of subterm positions (in particular positions filled with
variables) of $t$ with respect to $\sigma$?

\begin{example}\label{list-ex}
Let $\mathtt{List}/1$ and $\mathtt{Anylist}/0$ be type constructors, where 
$\tt List(\tau) \leq Anylist$ for all $\tau$, and $\mathtt{List}$ is the
usual list type, containing functions 
$\mathtt{Nil}_{\rightarrow \tt List(u) }$
and $\mathtt{Cons}_{\tt u, List(u)\rightarrow List(u)}$.
Consider the term
$\tt [x,[y]]$ (in usual list 
notation) depicted in Figure~\ref{list-fig}, and let
$\tt \sigma = Anylist$. Each functor in this term is introduced by an
application of Rule {\em (Func)}. Consider for example the term 
$\tt Nil$ in position $2.1.2$. Any type of it is necessarily an
instance of $\mathtt{List}(\mathtt{u}^{2.1.2})$, its declared type.\footnote{
We use the positions as superscripts to parameters in order to 
obtain fresh copies of those for every application of a rule.}
In order to derive that $\tt Cons(y,Nil)$ is a typed term, this
instance must be smaller (by the subtype order) than some instance of
the second declared argument type of $\tt Cons$ in position $2.1$,
that is, $\mathtt{List}(\mathtt{u}^{2.1})$.

\setlength{\unitlength}{0.4cm}
\begin{figure}[t]
\begin{center}
\begin{picture}(22,14)
\put(0,0){\framebox(3,2){$\tt y$}}
\put(3,0){\framebox(5,1){$\tt \mathtt{u}^\mathtt{y}$}}
\put(3,1){\framebox(5,1){${\tt u}^{2.1}$}}
\put(10,0){\framebox(3,2){$\tt Nil$}}
\put(13,0){\framebox(5,1){${\mathtt{List}(\mathtt{u}^{2.1.2})}$}}
\put(13,1){\framebox(5,1){${\mathtt{List}(\mathtt{u}^{2.1})}$}}
\put(5,2){\line(3,2){3}}
\put(13,2){\line(-3,2){3}}
\put(5,4){\framebox(3,2){$\tt Cons$}}
\put(8,4){\framebox(5,1){${\mathtt{List}(\mathtt{u}^{2.1})}$}}
\put(8,5){\framebox(5,1){$\mathtt{u}^2$}}
\put(15,4){\framebox(3,2){$\tt Nil$}}
\put(18,4){\framebox(5,1){$\mathtt{List}(\mathtt{u}^{2.2})$}}
\put(18,5){\framebox(5,1){$\mathtt{List}(\mathtt{u}^2)$}}
\put(10,6){\line(3,2){3}}
\put(18,6){\line(-3,2){3}}
\put(0,8){\framebox(3,2){$\tt x$}}
\put(3,8){\framebox(5,1){$\tt \mathtt{u}^\mathtt{x}$}}
\put(3,9){\framebox(5,1){$\tt \mathtt{u}^\epsilon$}}
\put(10,8){\framebox(3,2){$\tt Cons$}}
\put(13,8){\framebox(5,1){$\mathtt{List}(\mathtt{u}^2)$}}
\put(13,9){\framebox(5,1){$\mathtt{List}(\mathtt{u}^\epsilon)$}}
\put(5,10){\line(3,2){3}}
\put(13,10){\line(-3,2){3}}
\put(5,12){\framebox(3,2){$\tt Cons$}}
\put(8,12){\framebox(5,1){$\mathtt{List}(\mathtt{u}^\epsilon)$}}
\put(8,13){\framebox(5,1){$\tt Anylist$}}
\end{picture}
\end{center}
\caption{The term $\tt [x,[y]]$ and associated inequalities\label{list-fig}}
\end{figure}
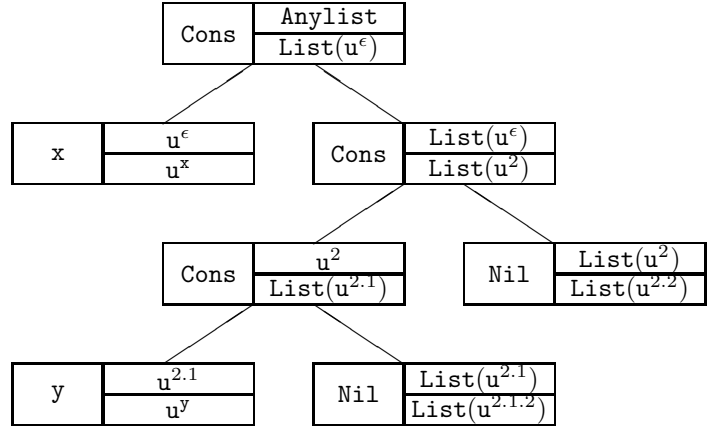

For the term in position $2.1.1$, the variable $\tt y$, a slightly
different consideration applies. Its type is given by a variable
typing. It is convenient to introduce a parameter $\tt \mathtt{u}^\mathtt{y}$ for this
variable and consider the type assigned to $\tt y$ by the variable
typing as an instance of $\tt \mathtt{u}^\mathtt{y}$. 

Analogous arguments can be applied to the
other subterms, and so in order to derive that 
$\tt [x,[y]]$ is a term of a type smaller than $\tt Anylist$, 
we are looking for an instantiation of
the parameters such that for each box corresponding to a position, the
type in the {\em lower} subbox is smaller than the type of the {\em upper}
subbox.  That is, we are looking for type substitutions 
such that 

\begin{eqnarray*}
\mathtt{u}^\mathtt{y}\Theta^\mathtt{y} &\leq&\mathtt{u}^{2.1}\Theta^{2.1}\\
\mathtt{List}(\mathtt{u}^{2.1.2})\Theta^{2.1.2} &\leq&\mathtt{List}(\mathtt{u}^{2.1})\Theta^{2.1}\\
\mathtt{List}(\mathtt{u}^{2.1})\Theta^{2.1} &\leq&\mathtt{u}^2\Theta^2\\
\mathtt{List}(\mathtt{u}^{2.2})\Theta^{2.2} &\leq&\mathtt{List}(\mathtt{u}^2)\Theta^2
\\
\mathtt{u}^\mathtt{x}\Theta^{\tt x} &\leq& \mathtt{u}^\epsilon\Theta^\epsilon\\
\mathtt{List}(\mathtt{u}^2)\Theta^2 &\leq&{\mathtt{List}(\mathtt{u}^\epsilon)}\Theta^\epsilon\\
\mathtt{List}(\mathtt{u}^\epsilon)\Theta^\epsilon&\leq&\mathtt{Anylist}
\end{eqnarray*}

\vspace{2ex}

For each position $\zeta$, the type substitution $\Theta^\zeta$
corresponds to the application of Rule {\em (Func)} that introduces
the functor in this position.  For each variable $x$, the type
substitution $\Theta^x$ defines a variable typing for $x$. Note
however that since the parameters in each application are renamed, we
can simply consider a single type substitution $\Theta$ which is the
union of all $\Theta^\zeta$.
\end{example}

We see that in order for $\_ \vdash t:\leq\sigma$
to hold, a solution
to a certain {\em type inequality system} (set of inequalities
between types) must exist. 

\begin{defi}
\label{inequality-def}
Let $t$ be a term and $\sigma$ a type such that 
$\_\vdash t:\leq\sigma$.
For each position $\zeta$ where $t$ has a
non-variable subterm, we denote the function in this position by
$f^\zeta_%
{\tau_1^\zeta,\dots,\tau_{n^\zeta}^\zeta\rightarrow\tau^\zeta}$
(assuming that the parameters in
$\tau_1^\zeta,\dots,\tau_{n^\zeta}^\zeta,\tau^\zeta$
are fresh, say by indexing them with $\zeta$).
For each variable $x$ occurring in $t$, we introduce a
parameter $u^x$ (so $u^x\not\in\pars(\sigma)$). 
The {\bf type inequality system} of $t$ and $\sigma$ is
\[
\begin{array}{ll}
{\mathcal I}(t,\sigma)=
\{\tau^\epsilon\leq\sigma\} \;\cup  &
\{\tau^{\zeta.i}\leq\tau^\zeta_i \mid
\mbox{Position $\zeta.i$ in $t$ is non-variable}\} \;\cup\\
 & 
\{u^x\leq \tau^\zeta_i \mid
\mbox{Position $\zeta.i$ in $t$ is variable $x$}\}.
\end{array}
\]
A {\bf solution} of ${\mathcal I}(t,\sigma)$ is a type substitution
$\Theta$ such that 
$\dom(\Theta)\cap\pars(\sigma)=\emptyset$ and 
for each $\tau\leq\tau'\in{\mathcal I}(t,\sigma)$, the inequality
$\tau\Theta\leq\tau'\Theta$ holds.

A solution $\Theta$ to ${\mathcal I}(t,\sigma)$ is {\bf principal}
if for every solution $\tilde{\Theta}$ for ${\mathcal I}(t,\sigma)$, 
there exists a $\Theta'$ such that for each 
$\tau\leq\tau'\in  {\mathcal I}(t,\sigma)$, we have
$\tau\tilde{\Theta}\leq\tau\Theta\Theta'$ and
$\tau'\tilde{\Theta}\leq\tau'\Theta\Theta'$.
\end{defi}

So for each subterm $f(\dots,g(\dots),\dots)$ of $t$, the type
inequality system says that the range type of $g$ must be less than or
equal to the $i$th argument type of $f$, where $g(\dots)$ is in the
$i$th position. 

If $\Theta$ is a solution for ${\mathcal I}(t,\sigma)$, 
by Prop.~\ref{inst-closed-prop}, for every type substitution 
$\Theta'$, we have that 
$\Theta \Theta'$ is also a solution for ${\mathcal I}(t,\sigma)$.
The following proposition follows from the rules in
Table \ref{rules-tab} and Def.~\ref{inequality-def}.

\begin{propo}\label{really-has-type}
Let $t$ be a term and $\sigma$ a type. If $U\vdash t:\leq\sigma$ 
for some variable typing $U$, then there exists a solution 
$\Theta$ for ${\mathcal I}(t,\sigma)$ (called the 
{\bf solution for ${\mathcal I}(t,\sigma)$ corresponding to $U$}) 
such that for each subterm 
$t'$ in position $\zeta$ in $t$, we have
\begin{itemize}
\item
$U\vdash t':\tau^\zeta \Theta$, if $t'$ is non-variable,
\item
$U\vdash t':u^x \Theta$, if $t'=x$ and $x\in\mathcal{V}$.
\end{itemize}
\end{propo}

The following lemma says that if $t$ is an instance of $s$, then a
solution to the type inequality system for $t$ is also a solution for
the type inequality system for $s$. 

\begin{rmlemma}\label{more-specific-lemma}
Consider two terms $s$ and $t$ such that 
$s$ is linear and
$s\theta=t$ for some idempotent $\theta$, and suppose that 
$\_\vdash s:\leq\sigma$ and
$\_\vdash t:\leq\sigma$. 
If $\Theta_t$ is a solution of ${\mathcal I}(t,\sigma)$, 
where 
$\dom(\Theta_t) \cap \pars({\mathcal I}(s,\sigma)) \subseteq \pars({\mathcal I}(t,\sigma))$, 
then 
\[
\tilde{\Theta}_s = 
\Theta_t \cup
\{u^x/\tau^{\zeta}\Theta_t \mid
\mbox{$s$ has $x$ in position $\zeta$ and $x\in\dom(\theta)$}\}
\]
is a solution of ${\mathcal I}(s,\sigma)$.
\end{rmlemma}
\begin{proof}
We first show that $\tilde{\Theta}_s$ is a well-defined type
substitution. Since $s$ is linear, $\zeta$ and hence
$\tau^{\zeta}\Theta_t$ is uniquely defined. Moreover, since $\theta$
is idempotent, $x$ cannot occur in $t$. Therefore 
$u^x\not\in \pars({\mathcal I}(t,\sigma))$, and hence by the condition 
on $\Theta_t$ in the statement, $u^x\not\in \dom(\Theta_t)$.

For the inequality $\tau^{\epsilon}\leq\sigma$ and for each 
$\tau^{\zeta.i}\leq\tau^{\zeta}_i\in {\mathcal I}(s,\sigma)$ such that 
$s$ has a non-variable term in $\zeta.i$, we have 
that the same inequality is also in 
${\mathcal I}(t,\sigma)$, and so $\Theta_t$, and consequently 
$\tilde{\Theta}_s$, is a solution for it.

For each 
$u^x\leq\tau^{\zeta}_i\in {\mathcal I}(s,\sigma)$ such 
that $x\in\dom(\theta)$, we have 
a corresponding inequality 
$\tau^{\zeta.i}\leq\tau^{\zeta}_i$ in ${\mathcal I}(t,\sigma)$.
Since 
$\tau^{\zeta.i}\Theta_t \leq
\tau^{\zeta}_i \Theta_t$ is true and
$\tau^{\zeta.i}\Theta_t = u^x \tilde{\Theta}_s$, it follows that
$u^x \tilde{\Theta}_s
\leq
\tau^{\zeta}_i \tilde{\Theta}_s$ is true.
\qed\end{proof}

\begin{example}\label{more-specific-example}
Let $s=\tt [x,z]$ and $t=\tt [x,[y]]$ and 
$\sigma=\tt Anylist$. A solution for 
${\mathcal I}(t,\sigma)$ is
\[
\Theta_t=
\{\mathtt{u}^\mathtt{y}/\mathtt{u}^{2.1}, 
\mathtt{u}^{2.1.2}/\mathtt{u}^{2.1},
\mathtt{u}^{\epsilon}/\mathtt{Anylist},
\mathtt{u}^{2.2}/\mathtt{Anylist}, 
\mathtt{u}^\mathtt{x}/\mathtt{Anylist},
\mathtt{u}^{2}/\mathtt{Anylist}\}
\]
(in Ex.~\ref{algo-ex} it will be shown how this solution is obtained).
Now 
\[
\begin{array}{rl}
{\mathcal I}(s,\sigma)=&
\{
\mathtt{u}^\mathtt{z}\leq \mathtt{u}^{2},
\mathtt{List}(\mathtt{u}^{2.2})\leq \mathtt{List}(\mathtt{u}^{2}),
\mathtt{u}^\mathtt{x}\leq \mathtt{u}^{\epsilon},
\mathtt{List}(\mathtt{u}^{2})\leq \mathtt{List}(\mathtt{u}^{\epsilon}),
\\&\;\:
\mathtt{List}(\mathtt{u}^{\epsilon}) \leq \mathtt{Anylist}
\}.
\end{array}
\] 
By Lemma~\ref{more-specific-lemma}, 
$\tilde{\Theta}_s = \Theta_t \cup
\{\mathtt{u}^\mathtt{z}/\mathtt{List}(\mathtt{u}^{2.1})\}$ is a solution for 
${\mathcal I}(s,\sigma)$. 
\end{example}

In the next subsection, we present an algorithm,
based on~\cite{FP98}, which computes a principal solution to a type
inequality system, provided $t$ is linear. In
Subsec.~\ref{max-var-typing-subsec}, our interest in principal
solutions will become clear. 

\subsection{Computing a Principal Solution}\label{algo-subsec}
The algorithm transforms the inequality system,
thereby computing bindings to parameters which constitute the
solution. It is convenient to consider system of both {\em
inequalities}, and {\em equations} of the form $u = \tau$. The
inequalities represent the current type inequality system, and the
equations represent the substitution accumulated so far. 
We use $\leqq$ for $\leq$ or $=$.

\begin{defi}\label{about-systems-def}
A system is {\bf left-linear} if each parameter occurs at most once on the
left hand side of an equation/inequality. 
A system is {\bf acyclic} if it does not have a subset
$\{\rho_1\leqq\sigma_1,...,\rho_n\leqq\sigma_n\}$ with
$\pars(\sigma_i)\cap\pars(\rho_{i+1})\neq\emptyset$ for all $1\le i\le n-1$, 
and
$pars(\sigma_n)\cap pars(\rho_1)\neq\emptyset$.
\end{defi}

\begin{propo}\label{special-form-prop}
If $t$ is a linear term, then any inequality system 
${\mathcal I}(t,\sigma)$ is acyclic and left-linear.
\end{propo}
\begin{proof}
Consider a non-variable position $\zeta$ in $t$.
There is exactly one inequality in
${\mathcal I}(t,\sigma)$ with $\tau^\zeta$ as left-hand side.  
Moreover, $\tau^\zeta$ is a flat type (declared range type of a
function), thus linear, and 
(because of indexing the parameters in $\tau^\zeta$ by $\zeta$) has no
parameters in common with any other left-hand side of 
${\mathcal I}(t,\sigma)$. 

Now consider a position $\zeta$ where $t$ has the variable $x$.
Because of the linearity of $t$, 
there is exactly one inequality in 
${\mathcal I}(t,\sigma)$ with $u^x$ as left-hand side.

Let $\{\rho_1\leq\sigma_1,...,\rho_n\leq\sigma_n\}$ be a subset of
${\mathcal I}(t,\sigma)$ with
$\pars(\sigma_i)\cap\pars(\rho_{i+1})\neq\emptyset$ for all 
$1\le i\le n-1$. 
By the definition of ${\mathcal I}(t,\sigma)$, if 
$\rho_1 = u^x$ for some variable $x$ or if
$\sigma_n = \sigma$, then 
$pars(\sigma_n)\cap pars(\rho_1)=\emptyset$.
If however 
$\rho_1\leq\sigma_1$ is 
$\tau^{\zeta.j}\leq\tau^\zeta_j$ and 
$\rho_n\leq\sigma_n$ is 
$\tau^{\xi.l}\leq\tau^\xi_l$ for some positions
$\zeta.j$ and $\xi.l$, 
then $\xi$ is a prefix of $\zeta$, 
and so, since we use the positions to index the parameters,
$pars(\sigma_n)\cap pars(\rho_1)=\emptyset$.
\qed\end{proof}

Example~\ref{list-ex} makes it also intuitively clear that
assuming linearity of $t$ is crucial for the above proposition.

We now give the algorithm for computing principal solutions
as a set of rules for simplifying a set of inequalities
and equations. A {\em solved form} is a system $I$ containing only equations
of the form $I=\{u_1=\tau_1,...,u_n=\tau_n\}$ where the parameters $u_i$ are
all different and have no other occurrence in $I$.
Note that the substitution $\{u_1/\tau_1,...,u_n/\tau_n\}$ associated to a solved form
is trivially a principal solution.

\begin{defi}\label{algorithm-def}
Given a type inequality system ${\mathcal I}(t,\sigma)$, where $t$
is linear, the {\bf type inequality algorithm} applies the following simplification rules:

\begin{tabular}{ll}
(1) & $\{K(\tau_1,...,\tau_m)\le K'(\tau'_1,...,\tau'_n)\}\cup I\longrightarrow
 \{\tau_{\iota (i)}\le \tau'_i\}_{i=1,..,n}\cup I$\\
& if $K\le K'$ and $\iota = \iota_{K,K'}$\\
(2)&$\{u\le u\}\cup I\longrightarrow I$\\
(3)& $\{u\le \tau\}\cup I\longrightarrow \{u=\tau\}\cup I[u/\tau]$\\
& if $\tau\not=u$, $u\not\in \vars(\tau)$.\\
(4) & $\{\tau\le u\}\cup I\longrightarrow \{u=\Max(\tau)\}\cup I[u/\Max(\tau)]$\\
& if $\tau\not\in V$, $u\not\in \vars(\Max(\tau))$ and
$u\not\in \vars(l)$ for any $l\le r\in \Sigma$.\\
\end{tabular}\\
\end{defi}

Intuitively, left-linearity of ${\mathcal I}(t,\sigma)$ is crucial
because it renders the binding of a parameter (point (3)) unique.

\begin{example}\label{algo-ex}
Consider ${\mathcal I}([x,[y]],{\tt Anylist})$ as in
Ex.~\ref{list-ex}. The initial $I$ is given by the inequality
system in the example (where the type substitutions are
removed). Applying (1) three times, we have
\[
I = \{
\mathtt{u}^\mathtt{y}\leq \mathtt{u}^{2.1}, 
\mathtt{u}^{2.1.2}\leq \mathtt{u}^{2.1},
\mathtt{List}(\mathtt{u}^{2.1})\leq \mathtt{u}^2,
\mathtt{u}^{2.2}\leq \mathtt{u}^2,
\mathtt{u}^\mathtt{x} \leq \mathtt{u}^\epsilon,
\mathtt{u}^2\leq \mathtt{u}^\epsilon\}.
\] 
Applying (3) five times, we have
\[
I = \{
\mathtt{u}^\mathtt{y} = \mathtt{u}^{2.1}, 
\mathtt{u}^{2.1.2} = \mathtt{u}^{2.1},
\mathtt{List}(\mathtt{u}^{2.1})\leq \mathtt{u}^\epsilon,
\mathtt{u}^{2.2} = \mathtt{u}^\epsilon,
\mathtt{u}^\mathtt{x} = \mathtt{u}^\epsilon,
\mathtt{u}^2 = \mathtt{u}^\epsilon\}.
\]
Applying (4) once, we have
\[
\begin{array}{rl}
I = &
\{
\mathtt{u}^\mathtt{y} = \mathtt{u}^{2.1}, 
\mathtt{u}^{2.1.2} = \mathtt{u}^{2.1},
\mathtt{u}^\epsilon = \mathtt{Anylist},
\mathtt{u}^{2.2} = \mathtt{Anylist},
\mathtt{u}^\mathtt{x} = \mathtt{Anylist},
\\&\;\:
\mathtt{u}^2 = \mathtt{Anylist}\}.
\end{array}
\]
\end{example}

\begin{propo}\label{alg-lemma}
Given a type inequality system ${\mathcal I}(t,\sigma)$, where $t$
is linear, the type inequality algorithm terminates
with either a solved form, in which case the 
associated substitution is a principal solution,
or a non-solved form in which case the system has no solution.
\end{propo}
\begin{proof}
Termination is proved by remarking that the sum of the sizes of the terms in left-hand sides
of inequalities strictly decreases after each application of a rule.

By Prop.~\ref{special-form-prop} the initial system is left-linear and acyclic,
and one can easily check that
each rule preserves the left-linearity as well as the acyclicity of the system.

Furthermore each rule preserves the satisfiability of the system
and its principal solution if one exists. Indeed rules (1) and (2) preserve all solutions
by definition of the subtyping order. 
Rule (3) replaces a parameter $u$ by its upper bound $\tau$.
As the system is left-linear this computes the principal solution for $u$,
and thus preserves the principal solution of the system if one exists. 
Rule (4) replaces a parameter $u$ having no occurrence in the left-hand
side of an inequality, hence having no upper bound,
by the maximum type of its lower bound $\tau$;
this computes the principal solution for $u$ and thus
preserves the principal solution of the system if it exists.

Now consider a normal form $I'$ for $I$. If $I'$ contains a non variable pair
$\tau\le\tau'$ irreducible by (1), then $I'$, and hence $I$, have no solution.
Similarly $I'$ has no solution if it contains an inequality
$u\le\tau$ with $u\in \vars(\tau)$ 
or an inequality $\tau\le u$ with $u\in \vars(\Max(\tau))$ (Prop.~\ref{corfini}).
In the other cases, 
by irreducibility and acyclicity, $I'$ contains no inequality,
hence $I'$ is in solved form and the substitution associated to $I'$
is a principal solution for $I$.
\qed\end{proof}

The next lemma says that principality is stable
under instantiation of types.

\begin{rmlemma}\label{max-closed-prop}
Let ${\mathcal I}(t,\sigma)$ be a type inequality system, where $t$ is
linear, and $\Theta'$
a type substitution such that 
$\dom(\Theta')\subseteq\pars(\sigma)$ and
$\ran(\Theta')\cap \pars({\mathcal I}(t,\sigma)) = \emptyset$. 
If  $\Theta$ is a principal solution of ${\mathcal I}(t,\sigma)$, 
then $\Theta\Theta'$ is a principal solution 
of ${\mathcal I}(t,\sigma\Theta')$.
\end{rmlemma}
\begin{proof}
Suppose $\Theta$ is computed by the algorithm of
Def.~\ref{algorithm-def}, and that $I_1,\dots,I_m$ is the sequence of
systems of this computation, i.e.~$\Theta$ is equal to $I_m$
viewed as a substitution. By Def.~\ref{inequality-def},
$\dom(\Theta)\cap\pars(\sigma)=\emptyset$.  In particular, this means
that no system $I_j$ ($j\in\onetom$) contains an inequality
$\tau\leq u$ where $u\in\pars(\sigma)$ and 
$\tau$ is not a parameter. It is easy to see that
$I_1\Theta',\dots,I_m\Theta'$ is a computation of the algorithm for
${\mathcal I}(t,\sigma\Theta')$, and hence $\Theta\Theta'$ 
(i.e.~$I_m\Theta'$ viewed as a substitution) is a principal solution
of ${\mathcal I}(t,\sigma\Theta')$.
\qed\end{proof}

\subsection{Principal Variable Typings}\label{max-var-typing-subsec}
The existence of a principal solution $\Theta$ of a type inequality system 
${\mathcal I}(t,\sigma)$ and Prop.~\ref{really-has-type} motivate 
defining the variable typing $U$ such that $\Theta$ is exactly the
solution of ${\mathcal I}(t,\sigma)$ corresponding to $U$.

\begin{defi}\label{max-vartyping-def}
Let $\_\vdash t:\leq \sigma$, and
$\Theta$ be a principal solution of ${\mathcal I}(t,\sigma)$.
A variable typing $U$ is {\bf principal} for $t$ and $\sigma$ if 
$U\supseteq
\{x:u^x\Theta \mid x\in\vars(t)\}$. 
\end{defi}

By the definition of a principal solution of
${\mathcal I}(t,\sigma)$ and Prop.~\ref{really-has-type}, 
if $U$ is a principal variable typing for $t$ and $\sigma$,
then for any $U'$ such that $U'(x) > U(x)$ for some $x\in\vars(t)$, 
we have $U'\not\vdash t:\leq\sigma$.
(since $U'$ corresponds to an instantiation of
the $u^x$'s that is not a solution of ${\mathcal I}(t,\sigma)$). 
The following is a corollary of Lemma~\ref{max-closed-prop}.

\begin{coro}\label{max-closed-coro}
If $U$ is a principal variable typing for 
$t$ and $\sigma$, then $U\Theta$ is a principal variable typing for 
$t$ and $\sigma\Theta$. 
\end{coro}

The following key lemma states conditions under which a substitution
obtained by unifying two terms is indeed ordered.

\begin{rmlemma}\label{is-ordered-substitution-lemma}
Let $s$ and $t$ be 
terms, $s$ linear, such that 
$U\vdash s:\leq\rho$, 
$U\vdash t:\leq \rho$, 
and there exists a substitution $\theta$ such that $s\theta = t$.
Suppose $\theta$ is a minimal matcher, i.e.~$\dom(\theta) \subseteq \vars(s)$.
Suppose 
$U$ is principal for $s$ and $\rho$. 
Then there exists a type substitution $\Theta$ such that for
$
U' = U\Theta\restr{\vars(s)} \,\cup\,
 U\restr{{\mathcal V}\setminus\vars(s)} 
$, we have that
$(\theta,U')$ is an ordered substitution. 
\end{rmlemma}
\begin{proof}
Since $\theta$ is a minimal matcher, we have 
\[
\theta= \{x/t' \mid 
\mbox{$\exists \zeta. x$ is subterm of $s$ in $\zeta$, 
$t'$ is subterm of $t$ in $\zeta$}\}.
\]

It remains to be shown that there exists a type substitution 
$\Theta$ such that $(\theta,U')$ as defined above is an ordered
substitution. 
Let $\Theta_s$ be the solution of ${\mathcal I}(s,\rho)$ corresponding
to $U$, and $\Theta_t$ be the solution of ${\mathcal I}(t,\rho)$ 
corresponding to $U$ (see Prop.~\ref{really-has-type}). Note that since
$U$ is principal for $s$ and $\rho$, $\Theta_s$ is a principal solution. 
By Lemma~\ref{more-specific-lemma}, 
$\tilde{\Theta}_s = 
\Theta_t\cup
\{u^x/\tau^{\zeta}\Theta_t \mid
\mbox{$s$ has variable $x$ in position $\zeta$}\}$ 
is a solution of ${\mathcal I}(s,\rho)$, and moreover,
since $\Theta_s$ is a principal solution of 
${\mathcal I}(s,\rho)$, there exists a type substitution
$\Theta$ such that for each 
$\tau$ occurring (on a left-hand side or right-hand side) in 
${\mathcal I}(s,\rho)$, 
\begin{equation}\label{tilde_s_is_smaller}
\tau \tilde{\Theta}_s \leq \tau \Theta_s \Theta.
\end{equation} 

\noindent
In particular, let $x$ be a variable occurring in $s$ in position
$\zeta$, and let $t'$ be the subterm of $t$ in position
$\zeta$. By Prop.~\ref{really-has-type}, 
$U'\vdash t':\tau^{\zeta}\Theta_t$. 
By Def.\ \ref{max-vartyping-def}, 
$x/u^x\Theta_s \in U$, and so by Rule 
{\em (Var)}, 
$U' \vdash
x:u^x\Theta_s \Theta$. Since by definition of $\tilde{\Theta}_s$, 
$\tau^{\zeta}\Theta_t = u^x \tilde{\Theta}_s$, we also have
$U' \vdash t':u^x \tilde{\Theta}_s$, 
and so by (\ref{tilde_s_is_smaller}), the condition in
Def.~\ref{ordered-subst-def} is fulfilled.
Since the choice of 
$x$ was arbitrary, the result follows.
\qed\end{proof}

\begin{example}\label{principal-typing-ex}
Consider the term vectors (since
Lemma~\ref{is-ordered-substitution-lemma} generalises in the obvious
way to term vectors)  
$s = \tt (3,x)$ and $t= \tt (3,6)$, let 
$\rho = \tt (Int,Int)$ and 
$U_s = \{\tt x:Int\}$, $U_t= \emptyset$ (see Ex.~\ref{sqrt-logic-ex}).
Note that $U_s$ is principal for $s$ and $\rho$, and so
$(\{{\tt x/6}\},U_s\cup U_t)$ is an ordered substitution ($\Theta$ is
empty).

In contrast, let $s = \tt (6,x)$ and $t= \tt (6,2.449)$, let 
$\rho = \tt (Real,Real)$ and 
$U_s = \{\tt x:Int\}$, $U_t= \emptyset$. 
Then $U_s$ is not principal for $s$ and $\rho$ (the principal variable
typing would be $\{\tt x/Real\}$), and indeed, there exists no
$\Theta$ such that $(\{{\tt x/2.449}\},U_s\Theta\cup U_t)$ is an ordered
substitution. 
\end{example}

\section{Nicely Typed Programs}\label{nicely-typed-sec}
In the previous section, we have seen that {\em matching}, 
{\em linearity}, and {\em principal variable typings} are crucial to
ensure that unification yields ordered substitutions (see
Lemma~\ref{is-ordered-substitution-lemma}). In this section, we define
three corresponding conditions on programs and the execution model.

We will generalise concepts defined for
{\em terms} in the previous section, to term {\em vectors}. In
particular, we consider principal variable typings for a
term vector $\vect{t}$ and a type vector $\vect{\sigma}$
(Def.~\ref{max-vartyping-def}). Also,
Lemma~\ref{is-ordered-substitution-lemma} generalises to term vectors
in the obvious way (conceptually, one could think of introducing
special functors into the typed language so that any vector can be
represented as an ordinary term).

First, we define modes, which are a common concept used for
verification~\cite{A97}.
For a predicate $p/n$, a {\bf mode} is an
atom $p(m_1,\dots,m_n)$, where 
$m_i \in \{{\tt \inp,\out} \}$ for $i \in \oneton$.
Positions with $\inp$ are called {\bf  input  positions}, and 
positions with $\out$ are called {\bf  output positions} of $p$.
We assume that a fixed mode is associated with each predicate in a program.
To simplify the notation, an atom written as $p(\vect{s},\vect{t})$ means: 
$\vect{s}$ is the vector of terms filling the  input positions, and
$\vect{t}$ is the vector of terms filling the output positions. 

\begin{defi}\label{moded-unification-def}
Consider a derivation step where $p(\vect{s},\vect{t})$ is the
selected atom and $p(\vect{w},\vect{v})$ is the renamed apart clause
head. 
The equation $p(\vect{s},\vect{t})=p(\vect{w},\vect{v})$ is 
{\bf solvable by moded unification} if there exist substitutions
$\theta_1$, $\theta_2$ such that 
$\vect{w}\theta_1=\vect{s}$ and
$\vars(\vect{t}\theta_1) \cap \vars(\vect{v}\theta_1) = \emptyset$
and 
$\vect{t}\theta_1\theta_2=\vect{v}\theta_1$.

A derivation where all unifications are solvable by moded unification is
a {\bf moded} derivation.
\end{defi}

Moded unification is a special case of {\em double matching}.
How moded derivations are ensured is not our problem here,
and we refer to~\cite{AE93}. Note that the requirement of moded
derivations is stronger than {\em input-consuming
derivations}~\cite{S99} where it is only required that the MGU does
not bind $\vect{s}$.

\begin{defi} \label{nicely-moded-def}
A query $Q =
p_1(\vect{s}_1,\vect{t}_1),\dots,p_n(\vect{s}_n,\vect{t}_n)$ 
is 
{\bf nice\-ly moded} if $\vect{t}_1,\dots,\vect{t}_n$ is a linear
 vector of terms and for all $i \in \oneton$
\begin{equation}
\vars(\vect{s}_i) \cap \bigcup_{j = i}^{n} \vars(\vect{t}_j)
= \emptyset. \label{nm-eq}
\end{equation}
The clause 
$C = p(\vect{t}_0,\vect{s}_{n+1}) \leftarrow Q$
is {\bf nicely moded} if $Q$ is nicely moded and
\begin{equation}
\vars(\vect{t}_0) \cap 
\bigcup_{j=1}^{n} \vars(\vect{t}_j) = \emptyset.
\label{head-eq}
\end{equation}
A program is {\bf nicely moded} if all of its clauses are nicely moded. 

An atom $p(\vect{s},\vect{t})$ is {\bf input-linear} if $\vect{s}$ is
linear, {\bf output-linear} if $\vect{t}$ is linear.
\end{defi}

\begin{defi}\label{nicely-typed-def}
Let 
\[
C = 
p_{\vect{\tau}_0,\vect{\sigma}_{n+1}}
(\vect{t}_0,\vect{s}_{n+1}) \leftarrow 
p^1_{\vect{\sigma}_1,\vect{\tau}_1}
(\vect{s}_1,\vect{t}_1),
\dots,
p^n_{\vect{\sigma}_n,\vect{\tau}_n}
(\vect{s}_n,\vect{t}_n)
\] 
be a clause. If $C$ is nicely moded, 
$\vect{t}_0$ is input-linear, and there exists a 
variable typing $U$ such that 
$U\vdash C\ \mathit{Clause}$, and
for each $i\in\zeroton$, 
$U$ is principal for $\vect{t}_i$ and $\vect{\tau}'_i$, where 
$\vect{\tau}'_i$ is the instance of $\vect{\tau}_i$ used for deriving 
$U\vdash C\ \mathit{Clause}$,
then we say that 
$C$ is {\bf nicely typed}.

A query $U_Q : Q$ is {\bf nicely typed} if the clause 
${\tt Go} \leftarrow Q$ is nicely typed.
A program is {\bf nicely typed} if all of its clauses are nicely typed.
\end{defi}

We can now state the main result.

\begin{rmtheorem}[Subject reduction]\label{persistence-thm}
Let $C$ and $Q$ be a nicely typed clause and query. If $Q'$ is
a resolvent of $C$ and $Q$ where the unification of the selected atom
and the clause head is 
solvable by moded unification, then $Q'$ is
nicely typed.
\end{rmtheorem}
\begin{proof}
By \cite[Lemma 11]{AL95}, 
$Q'$ is nicely moded.
Let $U_C$ and $U_Q$ be the variable typings used to type $C$ and $Q$,
respectively (in the sense of Def.\ \ref{nicely-typed-def}).

Let 
$p_{\vect{\sigma},\vect{\tau} }
(\vect{s},\vect{t})\in Q$ 
be the selected atom and
$C= p(\vect{w},\vect{v}) \leftarrow \mathbf{B}$.
By Rule {\em (Headatom)}, 
$U_C\vdash (\vect{w},\vect{v}):\leq(\vect{\sigma},\vect{\tau})$. 
Moreover, 
$U_Q\vdash (\vect{s},\vect{t}):\leq(\vect{\sigma},\vect{\tau})\Theta$
for some type substitution $\Theta$. Let 
$U = U_Q \cup U_C \Theta$. Note that since 
$\vars(C)\cap\vars(Q)=\emptyset$, $U$ is a
variable typing.
By Lemma \ref{variable-typing-lemma}, we have
$U\vdash \mathbf{B}\ Query$ and 
$U\vdash p(\vect{w},\vect{v})\ \mathit{Atom}$
(but not necessarily $U\vdash C\ \mathit{Clause}$, because of the
special rule for head atoms)
and in particular,
$U\vdash (\vect{w},\vect{v}):\leq
(\vect{\sigma},\vect{\tau})\Theta$.

Since $C$ is nicely typed, it follows by 
Cor.~\ref{max-closed-coro} that $U$ is principal for 
$\vect{w}$ and $\vect{\sigma}\Theta$. 
Moreover by assumption of moded unification, there exists a 
substitution $\theta_1$ such that
$\vect{w}\theta_1=\vect{s}$.
We assume $\theta_1$ is {\em minimal}, i.e.\ 
$\dom(\theta_1) \subseteq \vars(\vect{w})$.
By Lemma
\ref{is-ordered-substitution-lemma}, there exists a variable typing
$U'$ such that $(\theta_1, U')$ is an ordered
substitution, and moreover
$U'\restr{{\mathcal V}\setminus\vars(\vect{w})}= 
U\restr{{\mathcal V}\setminus\vars(\vect{w})}$. 
Therefore by Lemma \ref{apply-substitution-lemma}, 
$U' \vdash \mathbf{B}\theta_1\ Query$ and
$U' \vdash Q\theta_1\ Query$.
In particular, 
$U'\vdash \vect{v}\theta_1:\leq\vect{\tau} \Theta$.

Now since $Q$ is nicely typed and 
$U'\restr{\vars(Q)} = U_Q\restr{\vars(Q)}$, 
$U'$ is principal for $\vect{t}$ and $\vect{\tau}\Theta$.
Moreover by assumption of moded unification, there exists a minimal substitution
$\theta_2$ such that $\vect{t}\theta_2=\vect{v}\theta_1$. By Lemma
\ref{is-ordered-substitution-lemma}, there exists a variable typing
$U''$ such that
$(\theta_2, U'')$ is an ordered
substitution, and moreover 
$U''\restr{{\mathcal V}\setminus\vars(\vect{t})}= 
U'\restr{{\mathcal V}\setminus\vars(\vect{t})}$. 
Therefore by Lemma \ref{apply-substitution-lemma}, 
$U''\vdash \mathbf{B}\theta_1\theta_2\ Query$ and 
$U'' \vdash Q\theta_1\theta_2\ Query$.
Hence by Rule {\em (Query)}, 
$U''\vdash Q'\ Query$.
Finally, 
$U''\restr{{\mathcal V}\setminus(\vars(\vect{w})\cup\vars(\vect{t}))}=
U\restr{{\mathcal V}\setminus(\vars(\vect{w})\cup\vars(\vect{t}))}$ and so 
by the linearity conditions and (\ref{nm-eq}) in
Def.~\ref{nicely-moded-def}, it follows that
\begin{itemize}
\item
if $\vect{t}'$ is an output argument vector in $Q$, 
other than $\vect{t}$, 
and $\vect{\tau}'$ is the instance of the declared type of $\vect{t}'$
used for deriving $U_Q\vdash Q\ Query$,  
then $U'' \restr{\vars(\vect{t}')} = U_Q \restr{\vars(\vect{t}')}$, 
$\theta_1\theta_2 \restr{\vars(\vect{t}')} = \emptyset$, and hence 
$U''$ is a principal variable typing for 
$\vect{t}'\theta_1\theta_2$ and $\vect{\tau}'$, 
\item
analogously, if $\vect{t}'$ is an output argument vector in $\mathbf{B}$, 
and $\vect{\tau}'$ is the instance of the declared type of
$\vect{t}'$ used for deriving $U_C\vdash C\ \mathit{Clause}$, then 
$U'' \restr{\vars(\vect{t}')} = 
U_C\Theta \restr{\vars(\vect{t}')}$, 
$\theta_1\theta_2 \restr{\vars(\vect{t}')} = \emptyset$, 
and hence, by Cor.~\ref{max-closed-coro}, $U''$ is a principal
variable typing for $\vect{t}'\theta_1\theta_2$ and 
$\vect{\tau}'\Theta$.
\end{itemize}

So we have shown that $Q'$ is nicely moded, 
$U''$ is a variable typing such that 
$U''\vdash Q'\ Query$, and the principality
requirement on $U''$ is fulfilled. 
Thus $Q'$ is a nicely-typed query.
\qed\end{proof}

To conclude, we state subject reduction as a property of an entire derivation.

\begin{coro}
Any derivation for a nicely typed program $P$ and a nicely typed query
$Q$ contains only nicely typed queries.
\end{coro}

\begin{example}
Consider again Ex.~\ref{sqrt-logic-ex}.  The program is nicely typed, 
where the declared types are given in
that example, and the first position of each predicate is input, and
the second output. Both queries are nicely moded. The first query is
also nicely typed, whereas the second is not (see also
Ex.~\ref{principal-typing-ex}). For the first query, we have subject
reduction, for the second we do not have subject reduction.
\end{example}

\section{Discussion}\label{discussion-sec}\label{rel-work-sec}
In this paper, we have proposed criteria for ensuring subject
reduction for typed logic programs with subtyping
under the untyped execution model. Our starting point
was a comparison between functional and logic programming: In
functional programs, there is a clear notion of dataflow, whereas in
logic programming, there is no such notion a priori, and arguments can
serve as input arguments and output arguments. This difference is the
source of the difficulty of ensuring subject reduction for logic
programs. We thus coped with the problem by introducing modes into a
program, so that there is a fixed direction of dataflow. 

To understand better the numerous conditions for ensuring subject
reduction, it is useful to distinguish roughly between four kinds of conditions:
(1) ``basic'' type conditions on the program (Sec.~\ref{type-system-sec}), 
(2) conditions on the execution model (Def.~\ref{moded-unification-def}),
(3) mode conditions on the program (Def.~\ref{nicely-moded-def}),
(4) ``additional'' type conditions on the program
(Def.~\ref{nicely-typed-def}). We will refer to this distinction below.

Concerning (1), our notion of subtyping deserves discussion.
Approaches differ with respect to conditions on the 
{\em arities} of type constructors for which there is a subtype
relation. Beierle~\cite{B95} assumes that the (constructor) order is
only defined for type constants, i.e.~constructors of arity $0$. Thus
we could have $\tt Int \leq Real$, and so by extension 
$\tt List(Int) \leq List(Real)$, but not $\tt List(Int) \leq
Tree(Real)$.
Many authors assume that only constructors of the same arity are 
comparable. Thus we could have 
$\tt List(Int) \leq Tree(Real)$, but not $\tt List(Int) \leq Anylist$.
We assume, as~\cite{FP98}, that if $K\leq K'$, then the arity of $K'$ 
must not be greater that the arity of $K$. Other authors have been
vague about justifying their choice, suggesting that one could easily
consider modifications. 
We think that this choice is crucial
for the existence of principal types. In
particular, if one allowed for comparing constructors of arbitrary
arities, then the existence of a maximum above any type 
(Prop.~\ref{max-exists}) would not be guaranteed.

The PAN type system has been proposed in~\cite{MGS96} and described in
detail in~\cite{SG95}. It is argued there that 
comparisons between constructors of arbitrary arity should be allowed
in principle, and that the subtype relation should be defined by a
relation between argument positions of constructors, similar to our
$\iota$ (see Table \ref{subtyping-tab}). However, we believe that this
construction is flawed: It is claimed that under some simple
conditions, the subtyping relation implies a {\em subset} relation
between the sets of terms represented by the types, while in fact,
their formalism would allow for 
$\tt NonemptyList(Int) \leq List(String)$ (where those types are
declared as expected) even though the set of non-empty integer lists
is not a subset of the set of string lists. They define 
{\em extensional type bases}, essentially meaning typed languages
where also the converse holds, i.e., the subtyping relation exactly
corresponds to the subtype relation. 
Nothing is said about
decidability of this property, although the formalism heavily relies
on this concept.
Furthermore the very example given in order to motivate the need
for such a general subtyping relation is {\em not} extensional. 

Technically, what is crucial for subject reduction is that
substitutions are {\em ordered}: each variable is replaced with a term
of a smaller type. In Section~\ref{max-types-sec}, we give conditions
under which unification of two terms yields an ordered substitution:
the unification is a matching, the term that is being
instantiated is linear and is typed using a {\em principal}
variable typing. The linearity requirement ensures that a principle
variable typing exists and can be computed (Subsec.~\ref{algo-subsec}).
The conditions guarantee that the type of
each variable $x$ that is being bound to $t$ can be instantiated 
so that it is greater than the type of $t$. 

In Sec.~\ref{nicely-typed-sec}, we show how those conditions on the
level of a single unification translate to conditions on the program
and the execution model (points 2--4 above). We introduce modes and
assume that programs are executed using moded unification (2). This
might be explicitly enforced by the compiler by modifying the
unification procedure (which would have to yield a runtime error if 
the atoms are unifiable but violating the mode requirement). 
Alternatively, it can be verified statically that a program will be
executed using moded unification. In particular, nicely moded programs 
are very amenable to such  verification~\cite{AE93}. Moded unification 
can actually be very beneficial for efficiency, as witnessed by the 
language Mercury~\cite{mercury}. Apart from that, 
(3) nicely-modedness states the linearity of the terms being
instantiated in a unification. Nicely-modedness is designed so that it 
is persistent under resolution steps, provided clause heads are
input-linear. Finally, (4) nicely-typedness
states that the instantiated terms must be typed using a principal
variable typing.

Nicely-modedness has been widely used for verification purposes 
(e.g.~\cite{AE93}). In particular, the linearity condition on the
output arguments is natural: it states that every piece of data
has at most one producer. Input-linearity of clause heads however can
sometimes be a demanding condition, since it rules out equality tests
between input arguments~\cite[Section 10.2]{smaus-thesis}. 

Note that introducing modes into logic programming does
not mean that logic programs become functional. The aspect of
non-determinacy (possibility of computing several solutions for a
query) remains. 

Even though our result on subject reduction means that it is possible
to execute programs without maintaining the types at runtime, there
are circumstances where keeping the types at runtime is desirable, for 
example for memory management or for some extra logical operations like printing,
or in higher-order logic programming where the
existence and shape of unifiers depends on the types~\cite{NP92}.

There is a relationship between our notion of subtyping and {\em
transparency} (see Subsec.~\ref{transparency}). It has been observed in~\cite{HT92-new}
that transparency is essential for substitutions obtained from
unification to be typed. Transparency ensures that two terms of the
same type have identical types in all corresponding subterms, e.g. if
$\tt [1]$ and $\tt [x]$ are both of type $\tt List(Int)$, we are sure
that $\tt x$ is of type $\tt Int$. Now in a certain way, allowing for
a subtyping relation that ``forgets'' parameters undermines
transparency. For example, we can derive 
$\{{\tt x:String}\}\vdash [\mathtt{x}] = [1]\ \mathit{Atom}$, since $\tt List(String)
\leq Anylist$ and $\tt List(Int) \leq Anylist$, even though 
$\tt Int$ and $\tt String$ are incomparable. We compensate for this by
requiring principal variable typings. The principal variable typing 
for $\tt [x]$ and $\tt Anylist$ contains $\{\tt x:\mathtt{u}^\mathtt{x}\}$, and so 
$\tt \mathtt{u}^\mathtt{x}$ can be instantiated to $\tt Int$. However, our intuition is
that whenever this phenomenon (``forgetting'' parameters) occurs,
requiring principal variable typings is very demanding; but then, if
variable typings are not principal, subject reduction is likely to be
violated. 
As a topic for future work, we want to substantiate this intuition by
studying examples. In particular, we want to see if the conditions (in
particular, assuming principal variable typings) are {\em too}
demanding, in the sense that there are interesting programs that satisfy
subject reduction under more general assumptions. 

\subsection*{Acknowledgements}
We thank Erik Poll and Fran\c cois Pottier for interesting discussions on type systems for
functional programming, and the reviewers of the FSTTCS version of this article
for their valuable comments.
Jan-Georg Smaus was supported by an ERCIM fellowship.

\bibliography{thesis,modes_types,pierre}

 \newpage
\tableofcontents

\end{document}